\documentclass[twocolumn,showpacs,preprintnumbers,amsmath,amssymb]{revtex4}
\usepackage [dvips]{graphicx}

\begin{document}

\title{Effect of Deformation on Surface Characteristics of
Finite Metallic Crystals\footnote{Published in: Metallofiz.
Noveishie Tekhnol. 2002 -- \textbf{24} -- P.1651-1656 (in
parially); Ukr. Fiz. J. 2002 -- \textbf{47} -- P.1065-1071
(fully)}}

\author{V.~V.~ Pogosov\footnote{Corresponding author: E-mail:
 vpogosov@zstu.edu.ua (V.V.~Po\-go\-sov)}~ and~ O.~M.~ Shtepa}

\address{Department of Microelectronics, Zaporozhye National
Technical University, Zaporozhye 69064, Ukraine}

\date{\today}

\begin{abstract}
The   surface  stress  and  the  contact  potential  differences
of elastically  deformed  faces  of Al, Cu, Au,  Ni,  and  Ti
crystals  are calculated within the modified stabilized jellium
model using  the self -- consistent  Kohn--Sham method. The
obtained values of the  surface stress are  in  agreement  with
the  results of the  available first-principal calculations. We
find that the work function decreases/increases linearly with
elongation/compression of crystals. Our results  confirm that  the
available experimental data for the contact potential difference
obtained for  the deformed surface by the Kelvin method do not
correspond to  the change  of  the work function but to the change
of the surface potential. The problem of "anisotropy" of the work
function and ionization potential of finite sample is discussed.
\end{abstract}

\maketitle

\section{Introduction}

The  preparation and study of nanometer scale structures  attracts
a considerable  current  interest  for both  technological  and
scientific reasons. The free-electron gas models are invariably
popular tools in the physics of metals \cite{1,2} and
low--dimensional structures \cite{3}.

Recently,  several authors discussed the definition  of  the
surface stress  and the chemical potential for the case of finite
samples \cite{4,5,6,7,8}.

In our previous papers \cite{6,8} we proposed a successful method
to calculate the surface stress and the work function of a single
elastically deformed finite  metal crystal. It was applied for
simple metal. A fairly tightly bound  "d" band that overlaps and
hybridizes with a broader nearly free -- electron   "sp"  band
characterizes  transition  metals.  The cohesive properties of
simple, noble and transition metals can be calculated  from the
first principles in the context of one--electron picture provided
by the density-functional theory \cite{9,10}.

\begin{figure}[!t!b!p]
\centering
\includegraphics [width=0.5\textwidth] {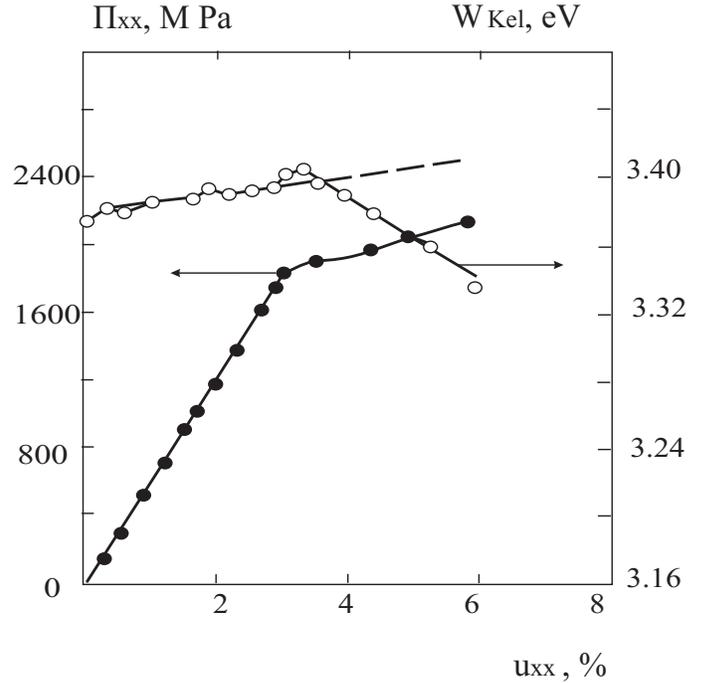}
\caption{The strain dependence of external mechanical stress
$\Pi_{xx}(u_{xx})$ and work function $W_{Kel}(u_{xx})$, defined by
Eq.(\ref{1}) for Al \cite{13}.} \label{Book-MS-5}
\end{figure}

\begin{figure}[!t!b!p]
\centering
\includegraphics [width=0.5\textwidth]{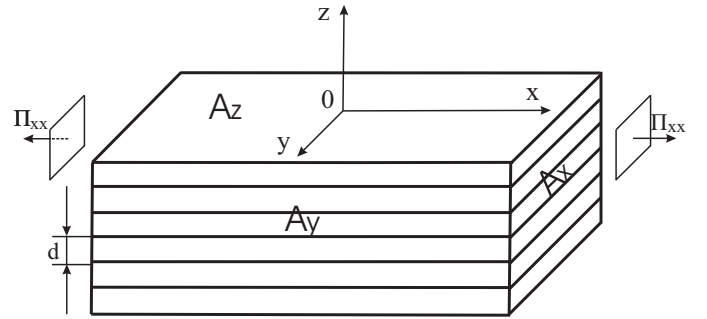}
\caption{Sketch of deformation.} \label{MS-1}
\end{figure}

The   direct   measurements,  using  the  Kelvin  method   showed
a decreasing/increasing of the contact potential difference  (CPD)
of  the elastically  tensed/compressed metal samples
\cite{11,12,13,14}. A similar effect  on CPD  was observed at the
surface of sample with a nonuniform distribution of  residual
mechanical stress \cite{15}. The conventional method of
measurement of the work function changes versus x--axes strain
\cite{11,12,13,14} is based on the expression:
\begin{equation}
\Delta W_{Kel} \equiv W(u_{xx})- W(0)= -CPD, \label{1}
\end{equation}
i.e.,  the work function as if increases for a tensed sample. The
change $\Delta W_{Kel}(u_{xx})$ was  measured (see
Fig.\ref{Book-MS-5}) at the side perpendicular y--  and
z--directions, $u_{xx}$ is the relative  deformation along x--axes
(see Fig.\ref{MS-1}). These,  at first
sight, surprising  results mean that the work function
increases/decreases  with uniaxial tension/compression of metallic
sample. This fact contradicts to another  well-known  observation:
the  work  function  of  simple metals decreases  for  the
transition Al $\Rightarrow$ Na $\Rightarrow$ Cs,  i.e.  with
decreasing of  the electron concentration.

In  the  present paper, we report the results of calculations of
the effect  of deformation on the surface energy, the work
function, and  the contact potential difference of faces of
technologically important metals such  as  Al, Ti, Ni, Cu, and Au
using their nominal valence. The problem of  an  accurate
definition of the work function is discussed. We checked the
accuracy of Eq.(\ref{1}) by fully self-consistent calculations and
showed that  it  is  wrong. Eq.(\ref{1}) is incorrect in the
diagnostic of strained surface.

\section{NONSTRAINED SURFACE}

Surface  energy  per  unit  area  and  work  function  are  the
most important characteristics of a metal surface. In the
framework of density--functional  theory  the  total energy of
metal is  a  functional of  the nonhomogeneous electron
concentration $n(r)$,  $n(r)\Rightarrow \bar{n}$ in the bulk,
$\bar{n}=3/4\pi r_{s0}^{3}$, $r_{s0}=Z^{1/3}r_{0}$, $r_{s0}$ and
$r_{0}$ are, respectively, the average between electron and
between ion distances in the nonstrained metal bulk, $Z$ is the
metallic valence. Using a pseudopotential approach the total
energy  can be written as the sum
\begin{equation}
E[n(r)]=T_{s}+E_{ex}+E_{cor}+E_{H}+E_{ps}+E_{M}, \label{2}
\end{equation}
where $T_{s}$ is the (non-interacting) electron kinetic energy,
$E_{ex}$ is the exchange energy, $E_{cor}$ is the correlation
energy, $E_{H}$ is the Hartree (electrostatic) energy, $E_{ps}$ is
the pseudopotential correction, and $E_{M}$ is the Madelung
energy. The  sum  of  first four terms in Eq.(\ref{2}) corresponds
to the energy  of "ordinary"  jellium, $E_{J}$. The "ordinary"
jellium model provides a useful description for the bulk modulus
and surface energy of the simple metals only  for $r_{s0}\approx $
4 bohr (Na), where bulk jellium is stable.  The jellium surface
energy for $r_{s0}\leq $ 2 and the bulk modulus for  $r_{s0}\geq$
5 bohr are negative. These deficiencies are  removed in the
stabilized jellium model \cite{17}. The average energy per valence
electron in the bulk of stabilized jellium is
$\bar{\varepsilon}_{SJ} =E[\bar{n}]/N$, where  $N$ is  a total
number of free electrons, defined by valence and atomic density,
\begin{equation}
\bar{\varepsilon}_{SJ}=\bar{\varepsilon}_{J}+\bar{w}_{R}+
\bar{\varepsilon}_{M}, \label{3}
\end{equation}
where the term ,
\begin{equation}
\bar{\varepsilon}_{J}=\bar{t}_{s}+\bar{\varepsilon}_{ex}+
\bar{\varepsilon}_{cor}=\frac{3}{10}\bar{k}_{F}^{2}+
\frac{3}{4\pi}\bar{k}_{F}+\bar{\varepsilon}_{cor}, \label{4}
\end{equation}
consists  of  the average kinetic,  exchange and correlation
energy  per electron,  $\bar{k}_{F}=(3\pi^{2} \bar{n}_{0})^{1/3}$,
$r_{0}$ is the radius of the Wigner -- Seitz cell, and $w_{R}=2\pi
\bar{n}_{0}r_{c}^{2}$ represents the average of the repulsive part
of the Ashcroft model potential, $r_{c}$ is the radius of ionic
core, $\bar{\varepsilon}_{M}=-9Z/10r_{0}$. We employ atomic units
throughout ($\hbar=m=e=1$) and the popular expression for the
correlation energy \cite{18,19}
\begin{equation}
\varepsilon_{cor}[n(r)]=
\frac{0.1423}{1+0.8293n(r)^{-1/6}+0.2068n(r)^{-1/3}}. \label{5}
\end{equation}

Applying  the  variation  principle one can find  the Euler --
Lagrange equation for nonhomogeneous electron gas
\begin{equation}
\mu=V_{eff}(r)+\frac{\delta T_{s}[n]}{\delta n(r)}, \label{6}
\end{equation}
where  $\mu$   is the chemical potential of the electrons, and
effective  one -- electron potential is
\begin{equation}
V_{eff}(r)=\varphi(r)+\frac{\delta [E_{ex}+E_{cor}]}{\delta
n(r)}+\langle\delta V\rangle_{face}\theta (r-r^{\prime}),
\label{7}
\end{equation}
where  $r^{\prime}$ is  the radius--vector of the surface, $\theta
(r-r^{\prime})$ is the step function.  The electrostatic potential
$\varphi(r)$ satisfies the Poisson equation.
\begin{equation}
\nabla^{2}\varphi(r)=-4\pi[n(r)-\rho(r)]. \label{8}
\end{equation}
The ionic charge distribution can be modelled by the step function
$\rho(r)=\bar{\rho}\theta (r-r^{\prime})$, where
$\bar{\rho}=\bar{n}_{0}/Z$.

The  coordinate--independent term $\langle\delta V\rangle_{face}$
in Eq.(\ref{7}) represents the difference between  the
pseudopotential of the ion lattice  and the electrostatic
potential  of  positive background averaged over the Wigner--Seitz
cell. This  term  allows  one to distinguish different faces  of
semi-infinite samples. The face-dependence of the stabilization
potential \cite{16,17} reads
\begin{equation}
\langle\delta V\rangle_{face}=\langle\delta
V\rangle_{WS}-\left(\frac{\bar{\varepsilon}_{M}}{3}+
\frac{\pi\bar{n}_{0}}{6}d_{0}^{2}\right), \label{9}
\end{equation}
\begin{equation}
\langle\delta V\rangle_{WS}=
\bar{n}_{0}\frac{d}{d\bar{n}_{0}}(\bar{\varepsilon}_{M}+
\bar{w}_{R}), \label{9a}
\end{equation}
where  $d_{0}$ is the spacing between the lattice planes parallel
to the surface. From the bulk stability condition
$d\bar{\varepsilon}_{SJ}/d\bar{n}_{0}=0$ one can obtained the
relation
\begin{equation}
\langle\delta V\rangle_{WS}=-
\bar{n}_{0}\frac{d}{d\bar{n}_{0}}(\bar{t}_{s}+\bar{\varepsilon}_{ex}+
\bar{\varepsilon}_{cor}), \label{10}
\end{equation}
The electronic profile $n(r)$ can be expressed in the terms of
one-electron wave functions $\psi_{i}$,
\begin{equation}
n(r)=\sum\limits_{i=1}^{N}|\psi_{i}(r)|^{2}, \label{11}
\end{equation}
where the functions $\psi_{i}$ satisfy the one-electron wave
equation
\begin{equation}
-\frac{1}{2}\nabla^{2}\psi_{i}(r)+V_{eff(face)}(r)\psi_{i}(r)=
\varepsilon_{i}\psi_{i}(r). \label{12}
\end{equation}
The set of the Kohn-Sham equations must be solved
self-consistently. Then the total kinetic energy of electrons in
Eq.\ref{2} is
\begin{equation}
T_{s}[n]=\sum\limits_{i=1}^{N}\varepsilon_{i}-\int
d^{3}rn(r)V_{eff}(r) \label{12a}
\end{equation}

As  a rule for semi-infinite metal it is supposed that the
electronic profile $n(r)$   and   effective   potential
$V_{eff}(r)$ vary only in the direction perpendicular   to  the
surface.  The conventional approach involves introducing periodic
boundary conditions in the x-- and y--directions. Thus, only
crystal face is in z--direction. In this case the set  of
equations (\ref{8}), (\ref{11}) and (\ref{12}) reduces to
\begin{equation}
\varphi(z)=
\varphi(\infty)-4\pi\int\limits_{z}^{\infty}dz^{\prime}
\int\limits_{z^{\prime}}^{\infty}dz^{\prime\prime}
[n(z^{\prime\prime})-\rho(z^{\prime\prime})],\label{13}
\end{equation}
\begin{equation}
n(z)=\frac{1}{\pi^{2}}\int\limits_{0}^{\bar{k}_{F}}dk
\left(\bar{k}_{F}^{2}-k^{2}\right) \left| \psi _{k}(z)\right|^{2},
\label{14}
\end{equation}
\begin{equation}
\left[-\frac{1}{2}\frac{d^{2}}{dz^{2}}+V_{eff}[z,n]\right]\psi_{k}(z)=
\frac{1}{2}k^{2}\psi_{k}(z). \label{15}
\end{equation}
Here the effective potential is taken in a
local-density-approximation
\begin{equation}
V_{eff}[z,n]=\phi (z)+V_{xc}(z)+\left\langle \delta V\right\rangle
_{face} \theta (-z), \label{16}
\end{equation}
where  $n\equiv n(z)$.  The  electrons wave number  varies in the
interval (0, $\bar{k}_{F}$).  The solution  of  the  problem can
be reduced to the iteration procedure  by means of the follow
relation \cite{20}
\begin{equation}
\phi_{i+1}(z)=\int\limits_{-\infty}^{\infty}dz^{\prime}
e^{-k_{F}|z-z^{\prime}|}\frac{2\pi}{k_{F}}[n(z^{\prime})-\rho(z^{\prime})]+
\frac{k_{F}}{2}\phi_{i}(z),\label{16a}
\end{equation}
where $i$ is the number of iteration.

The surface energy can be written as
\begin{equation}
\gamma_{SJ}=\gamma_{J}+\langle V_{WS}\rangle
\int\limits_{-\infty}^{0}dz[n(z)-\bar{n}_{0}], \label{17}
\end{equation}
The ordinary jellium components are:
\begin{multline}
\gamma_{s}=\frac{1}{2\pi^{2}}\int\limits_{0}^{k_{F}}dkk\left[\frac{\pi}{4}-
\delta (k)\right]-\\-\int\limits_{-\infty}^{\infty}dzn(z)\left[V_{
eff(face)}(z)- \bar{V}_{eff(face)}\theta(-z)\right],\label{18}
\end{multline}
where $\delta$ is the phase shift of wave function,$\psi
_{k}(z)\rightarrow \sin[kz-\delta(k)]$ as $z \rightarrow -\infty$,
$\bar{V}_{eff(face)}$ is the bulk magnitude of effective
potential, and
\begin{equation}
\gamma_{xc}=\int\limits_{-\infty}^{\infty}dz[n(z)\varepsilon_{xc}(n(z))
-\bar{n}\varepsilon_{xc}(\bar{n})\theta(-z)], \label{19}
\end{equation}
\begin{equation}
\gamma_{H}=\frac{1}{2}\int\limits_{-\infty}^{\infty}dz\varphi(z)
[n(z)-\rho(n(z))]. \label{20}
\end{equation}
Putting the electrostatic potential in the vacuum equal zero,
$\varphi(+\infty)$ =0, one can calculate the work function as
$(\bar{\varepsilon}_{F}=\bar{k}_{F}^{2}/2)$
\begin{equation}
W_{z-face}\equiv-\mu=-\bar{V}_{eff}-\bar{\varepsilon}_{F}.
\label{21}
\end{equation}

\section{DEFORMED SURFACE}

The  strain  dependence of the CPD was measured  for
polycrystalline compressed  \cite{11}  and  tensed  samples
\cite{13,14}. One can  assume that  a polycrystal  is  assembled
from a number of simple  crystallites. Thus, qualitatively, the
problem can be reduced to an analysis  of tension  or compression
applied to a single crystal. We consider a single crystal  in the
shape of  a parallelepiped  with sides  having  equivalent Miller
indices. For simplicity, the material of the sample is assumed to
have the cubic crystallographic symmetry. We introduce the
coordinate system with  axes  and  perpendicular to the sample
faces. Their areas are equal to   and respectively (see
Fig.\ref{MS-1}). We neglect temperature and dimensional effects
that differ from the approach used in \cite{3,21}.

Let  us  first express the average electron density in a metal  as
a function  of deformation. For this purpose, consider an
undeformed cubic cell  of  side length $a_{0}$, $a_{0}^{3}=4\pi
r_{0}^{3}/3$. In the modified stabilized jellium model \cite{6,8}
the metal energy is a function of the electron  density parameter
$r_{su}=Z^{1/3}r_{0u}$, spacing $d_{u}$ between the lattice planes
perpendicular to z--direction, the Poisson coefficient for
polycrystal $\nu$, the Young modulus $Y$, and deformation
$u_{xx}$.

For  uniaxially deformed cell elongated or compressed  along  the
x--axis one can write \cite{8}:
\begin{equation}
       d_{u}=d_{0}\left( 1-\nu u_{xx}\right),        \label{22}
\end{equation}
where  $d_{u}$ is the spacing between the lattice planes
perpendicular to the  y-- or z--directions, $d_{0}$ is the
interplanar spacing in undeformed crystal,
\begin{equation}
\bar{n}=\bar{n}_{0}\left[ 1-\left( 1-2\nu \right)u_{xx} \right]
+O\left( u_{xx}^{2}\right) , \label{23}
\end{equation}
is  the  average electron density in deformed metal bulk, $\nu$ is
the Poisson coefficient for polycrystal and the corresponding
density parameter is .
\begin{equation}
r_{su}=r_{s}\left[ 1+( 1-2\nu) u_{xx}\right] ^{1/3}. \label{24}
\end{equation}
The stabilization potential (\ref{10}) must be changed by
\begin{equation}
\langle\delta V\rangle_{WS}=-
\bar{n}_{0}\frac{d}{d\bar{n}_{0}}\left(\bar{t}_{s}+\bar{\varepsilon}_{ex}+
\bar{\varepsilon}_{cor}+\frac{P}{\bar{n}_{0}}\right), \label{25}
\end{equation}
where
$$
P=-(\Pi_{xx}+\Pi_{yy}+\Pi_{zz})=-Yu_{xx}(1-2\nu)
$$
is the external pressure.

An  applied  force  stimulates the change of a volume  and
reticular electron  density  at  particular faces of neutral
metallic crystal  and results in a difference between
electrostatic potentials of these faces.

The corresponding anisotropic three-dimensional electric field
arises due to a transfer of electrons between faces. Thus, finite
sample sizes lead, in   the   first   place,  to  the  fact  that
the  surface tends   to equipotentiality.  In general case, this
transfer of electrons does  not give  any  possibility  to
calculate the work function  using conventional approach  for
semi--infinite sample. However, in the special case of  the
largest face of the metallic crystal,
\begin{equation}
A_{z}\gg A_{x},A_{y}, \label{26}
\end{equation}
the  use of the conventional approach (see Eq. (\ref{21})) in fact
is adequate \cite{8}, $A_{z}$ is the area of upper side at Fig. 2.
As a result, the calculated work function exclusively  for
z--direction corresponds to that for the \emph{whole} crystal. The
face perpendicular to this direction  is not appreciably perturbed
by the  transferred electronic charge independently  from  the
crystallographic orientation \cite{8}. The conventional method
involves introducing periodic boundary conditions  in the  x-- and
y--directions.  In this case the surface  potential depends
strongly from the atomic packing density. Thus, the only crystal
face  is in  -direction. Putting the three-dimensional
electrostatic potential  in the vacuum equal zero, under condition
(\ref{26}) one can use definition (\ref{21}), where  the effective
potential in the bulk of semi-infinite metal yields the  total
sample-vacuum barrier. Solving the set of Eqs. (\ref{13}) --
(\ref{16}) and using relations (\ref{22}) -- (\ref{25}), one can
calculate the strain dependencies of the surface energy and work
function.

We  calculate  the  diagonal component of surface  stress  for
given largest face using the following expression \cite{5,6,7,8}:
\begin{equation}
\tau_{xx}=\gamma+\frac{d\gamma}{du_{xx}}, \label{27}
\end{equation}
where $\gamma$ is the surface energy per unit area of the largest
face of sample.

\section{RESULTS AND DISCUSSIONS}

We  performed our calculations for the work function and the
surface energy for the case of absence of the strain state of
metals and for  the strain dependences $W_{face}(u_{xx})$ and
$\gamma_{face}(u_{xx})$ within the range of deformation  ïðåäåëàõ
äåôîðìàöèé $-0.01 \le u_{xx} \le +0.01$ for Ni and $-0.03 \le
u_{xx} \le +0.03$ for Al,  Au, Cu, and Ti, respectively. The
positive/negative deformation   is equivalent to the
tension/compression of the largest side of the sample, i.e., the
decrease/increase of the atomic packing density at this side, and
the  decrease/increase of the mean electron concentration and the
interplanar  spacing  in direction. Upper side of sample  in Fig.\ref{MS-1}  is suggested as having indexes (100), (110), (111) or
(0001).

The  tends in bulk and surface characteristics of Al, Au, Cu, Ni,
and Ti  have been reproduced from the theory of uniform electron
gas with the average  electron  density corresponding to the
nominal  valence by  the volume  per  atom. The use of the integer
and fractional magnitude of  a valency  leads to a successful
calculation of bulk modulus  $B$ \cite{22}.  The concept  of  the
metallic valency can be defined  to  treat simple  and transition
metals in terms of the uniform electron gas.

\begin{figure}[!t!b!p]
\centering
\includegraphics [width=0.5\textwidth] {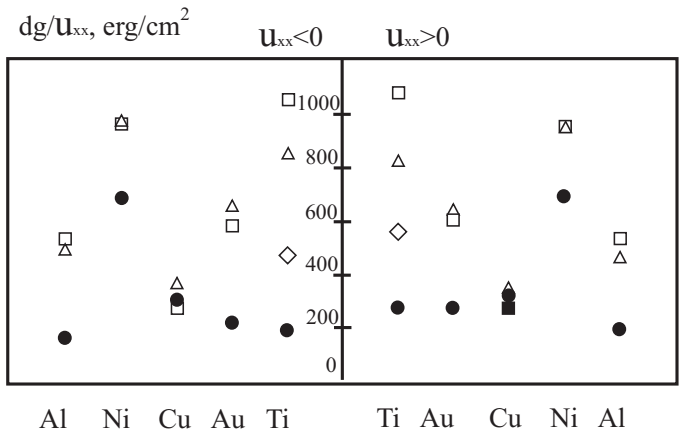}
\caption{Calculated derivative $d\gamma/du_{xx}$ for estimation of
the surface stress, Eq.(\ref{26}). The left and right parts of the
figure correspond to  the compression ($u_{xx}<0$) and the tension
($u_{xx}>0$) of the sample, respectively: $\Box$ -- fcc (100),
$\triangle$ -- fcc (110), $\bullet$ -- fcc (111), $\Diamond$ --
hcp (0001).} \label{Book-MS-7}
\end{figure}

The  results  of  our  calculations for free and deformed  faces
are summarized  in Table \ref{SH-2} and Figs.2 and 3,
respectively. For
\begin{table}
\centering \caption{Calculated bulk modulus $B$, surface energy
$\gamma_{face}$, and the work function  $W_{face}$. For univalent
Au ($Z=1$)the results are placed in brackets. The experimental
values  of  $B$ (are given), $\sigma$, $W_{face}$ for
polycrystaline metal, and the Poisson  ratio $\nu$, the Young
modulus  $Y$ necessary for calculation  the strain dependences are
taken from books \cite{23,24,25}. } \vspace*{0.5cm}
\smallskip
\tabcolsep .1cm \label{SH-2}
\begin{tabular}{|c|c|cccc|c|} \hline \hline
 &$B$ & Face & $\sigma_{face}$ &$W_{face}$ & $Y$ &$\nu$ \\
&[Mbar]&&[erg/cm$^2$]&[eV]&[GPa]&\\\hline\hline
   &           & (100) & 1087  & 3.806 & 62.5 & \\
Al     &1.565/0.722& (110) & 1683  & 3.643 & 71.4 & 0.34 \\
     &           & (111) & 939   & 4.119 & 75.1 &  \\\hline
   &           & (100) &  1069(395)  & 3.792(3.318) & 43.5 & \\
Au&1.519(0.202)/0.722& (110) & 1652(440)  & 3.630(3.148) & 81.3 & 0.42 \\
     &           & (111) & 924(383)   & 4.105(3.478) & 115.0 &  \\\hline
  &           & (100) & 1376  & 4.010 & 138.0 & \\
 Ni     &2.567/1.860& (110) & 2224  & 3.858 & 215.0 & 0.32 \\
     &           & (111) & 1162  & 4.325 & 262.0 &  \\\hline
   &           & (100) & 979  & 3.855 & 65.8 & \\
  Cu   &1.113/1.370& (110) & 1295 & 3.647 & 131.0 & 0.35 \\
     &           & (111) & 899  & 4.123 & 194.0 &  \\\hline
  &           & (0001)& 1081  & 4.205& 145.0 & \\
     &           & (100) & 1355  & 3.865 & 96.1 & \\
Ti      &1.565/0.722& (110) & 2456  & 3.774 & 96.1 & 0.30 \\
     &           & (111) & 1081  & 4.205 & 27.8 &  \\\hline
     \hline
\end{tabular}
\end{table}
comparison  we perform  calculations of bulk modulus
$B=\bar{n}_{0}^{2}d^{2}(\bar{n}_{0}^{2}\bar{\varepsilon}_{SJ})
/d\bar{n}_{0}^{2}$, $\gamma$ and $W$ for Au ($Z$ = 1). The data of
Table \ref{SH-2} demonstrate intricate picture for common
description of cohesive and  surface metallic properties in the
simple approach. In fact the  use of nominal valence gives
satisfactory  results for   the surface characteristics.  We found
that the deformation changes of the work function  and the surface
energy remain linear with respect to  the deformation. The strain
derivative of the work function and  the surface energy is
positive.  The values  of surface stress component vary
appreciably  within the interval (1.15, 1.75)$\gamma_{face}$.
Similar results for  the surface stress (1250 and 1440
erg/cm$^{2}$ for Al (111)) are yielded by the \emph{ab initio}
\cite{26} and atomistic \cite{27} calculations. In contrast to the
present work (Fig.\ref{Book-MS-7}), the strain derivative obtained
for Au in \cite{28} was larger than the surface energy.

\begin{figure}[!t!b!p]
\centering
\includegraphics [width=0.5\textwidth] {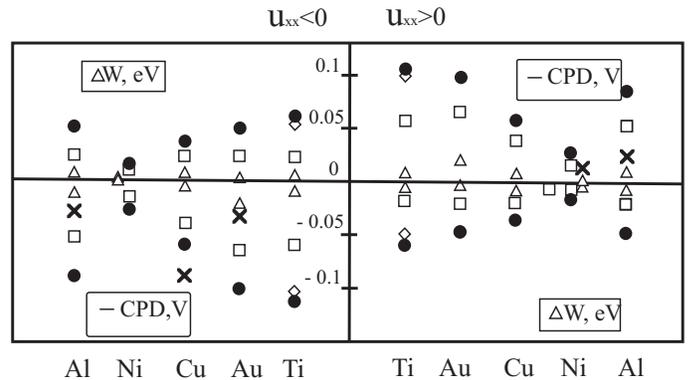}
\caption{Calculated  change  of  the work  function  and  the
contact potential  difference for maximal strain of elastically
deformed largest faces (Fig.\ref{MS-1}). The experimental values
of $-CPD$ taken from \cite{11} for compressed ($u_{xx}=-0.03$)
polycrystalline Al, Cu, and Au samples, and from \cite{13,14} for
tensed Al ($u_{xx}=+0.03$)  and  Ni ($u_{xx}=+0.01$) samples (see
also Table 1 in \cite{29}). The values of for negative deformation
are extracted from the lowest experimental shifts in contact
potential $-CPD$ (measured in units of $\mu$V$\times$cm$^{2}$/kg)
of \cite{11} multiplied by $\bar{Y}|u_{xx}|$, where $\bar{Y}=
\frac{1}{3}(Y_{001}+Y_{110}+Y_{111})$. Note that the values of and
were erroneously equated in \cite{11,12,13,14,15}.}
\label{Book-MS-6}
\end{figure}

The  relative change of the work function equals approximately 1\%
for the  maximal  strains (compare Table \ref{SH-2} and
Fig.\ref{Book-MS-6}). For the
case of  the compression  ($u_{xx}<0$),  the  tail of electron
profile and, therefore, of  the effective potential grows steeper
in vacuum. For the case of the tensile strain  ($u_{xx}>0$), these
coordinate dependences have the opposite tendency.  A total
decrease/increase of  the  work  function   is  determined by  a
positive/negative shift of the effective potential versus strain
in  the bulk  of metal (neglecting the deformation dependence
$\varepsilon_{F}(u_{xx})$, one can certainly put  the  shift
$\Delta W\simeq -\Delta \bar{V}_{eff}$). Our calculation  mimics
the usual work function dependence from electronic concentration
(for "transition" Al $\Longrightarrow$ Na $\Longrightarrow$ Cs).

However,  relation  (\ref{1}) gives incorrect dependence
$W(u_{xx})$. These results,  at first sight, contradict to
experiments \cite{11,12,13,14}.

We  suggest that, qualitatively, the problem is reduced to the
effect of  strain  on  a  single crystal. The experimental
observations can  be explained based on the change of the
effective potential at the position of the image plane $z=z_{0}$
\cite{8}. For simplicity we take  $z_{0}$ = 1 bohr for faces and
use
$$
CPD=\Delta V_{eff}(z_{0},u_{xx}).
$$
At  present,  we  calculated $\Delta W$ and $CPD$ without use of
Eq.(\ref{1}).  Our Fig.\ref{Book-MS-6} demonstrates,  on  the one
hand, a good qualitative agreement of  the calculated values of
with the experimental data, on the other hand,  the inverse
dependence than  it follows from  Eq.  (\ref{1}).

Let us introduce $\xi$ the ratio   of the effective potential
differences  between strained and strain-free samples at the
surface and in the bulk,
$$
\xi=\frac{\Delta V_{eff}(z_{0},u_{xx})}{\Delta
\bar{V}_{eff}(u_{xx})}\simeq\frac{CPD}{-\Delta W},
$$
where $\bar{V}_{eff}(u_{xx})\equiv V_{eff}(z=-\infty,u_{xx})$. The
calculations show that the value $\xi$  varies within the interval
(-3, -1) for the metal faces under study. For Al $\xi\simeq
-2.8$!!!

The analysis of experimental data provided the evidence that it is
not adequate to use the Kelvin method for the measurement, e.g.,
of temperature dependence of the work function (see Fig. 6 of
\cite{30}).

Our  results show:

(i) the strain changes of the effective potential in  the bulk and
at the surface have the \emph{opposite signs},

(ii) the sign of the deformation  effect  is  \emph{independent}
on  the material  and of  the crystallographic orientation.

In  the  conclusion  let  us  remember  that  the
density-functional calculations  of  this quantity deal with the
nonrealistic semi-infinite systems. However, in all experiments
\emph{finite} samples are used, in general, of  arbitrary  sizes
and shapes. In this context,  an  important remark appears   about
the conventional  definition  of  face--dependent work function.
The ionization  potential ($IP$) of a finite  sample  (as  a
"giant molecule") is given by
\begin{equation}
IP=E_{N-1}-E_{N}\simeq W+\frac{e^{2}}{2C}, \label{28}
\end{equation}
where $E_{N}$  is  the  total  energy  of  neutral  metallic
sample containing $N$ electrons. Here  $W$ is introduced as a work
function of the \emph{whole} sample and $C$ is a capacitance of
the sample, which is a scalar quantity. It is clearly seen  from
Eq.(\ref{28}) that  the work function is  not  an "anisotropic"
characteristic but a scalar quantity, and $IP\longrightarrow W$ at
$C\longrightarrow \infty$. It may be demonstrated  by formation of
finite sample adding consistently atoms each to other.  The
ionization  potential of single atom, dimer, etc. (up to  work
function) corresponds to the ionization process of each stage of
nucleation (up  to the  solid) \cite{31}. Thus, the conventional
definition (\ref{28})  leads to  the conclusion that  the concept
of an anisotropy of the  work function  of metal is spurious   in
principle.  Hence,  the realistic geometry considerations
presented in the \cite{16,32} are adequate in the case when  all
sample faces is posses the same atomic packing density.

Our theory can be applied for the explanation of new pressure
effects in a single-electron transistor  \cite{33}.

\acknowledgments{This  work was supported by the Ministry of
Education and Science of Ukraine and the NATO "Science for Peace"
Programme. We thank Prof. V. U. Nazarov for  giving  us an access
to his computer code for testing purposes  and Dr. W. V. Pogosov
for reading the manuscript.}


\end{document}